\documentstyle[12pt]{article}

\newcommand{\cgn}[6]{(#1,#2,#3,#4 \vert #5,#6)}
\newcommand{\cg}[9]{(#1,#2,#3,#4,#5,#6 \vert #7,#8,#9)}

\newcommand{\qpn}[1]{[ \! [#1] \!]_{qp}}

\newcommand{\Qn}[1]{[#1]_{Q}}

\newcommand{\udm}{\frac{1}{2}}

\def\gr{\hbox{\bf R}}

\def\grn{\hbox{\bf N}}
\def\grz{\hbox{\bf Z}}
\def\grc{\hbox{\bf C}}

\begin{document}

\vskip 4 cm 
\centerline {\bf APPLICATION OF A TWO-PARAMETER QUANTUM ALGEBRA} 

\centerline {\bf TO ROTATIONAL SPECTROSCOPY OF NUCLEI\footnote{This 
work was presented by one of the authors (R.~B.) 
to the symposium ``Quantum Groups and Their Applications in Physics'' 
(Pozna\'n, Poland, 17-20 October 1995). To be published in Reports on 
Mathematical Physics.}}
\vskip 2 cm
\centerline {R. BARBIER and M. KIBLER}
\vskip 1 cm
\centerline {Institut de Physique Nucl\'eaire de Lyon, 
IN2P3-CNRS et Universit\'e Claude Bernard,} 

\centerline {43 Boulevard du 11 novembre 1918, 
             F-69622 Villeurbanne Cedex, France}

\vskip 3.6 cm
\vskip 0.9 true cm

A two-parameter quantum algebra 
                                $U_{qp}({\rm u}_2)$ is briefly investigated in
this paper. The basic ingredients of a model based on the $U_{qp}({\rm u}_2)$ 
symmetry, the $qp$-rotator model, are presented in detail. Some general 
tendencies arising from the application of this model to the description of
rotational bands of various atomic nuclei are summarized. 

\newpage 


{\bf 0. Introduction}

\vskip 0.6 true cm 

In recent years, the theory of compact matrix quantum groups (or pseudogroups) 
\cite{Wor1,Wor2,Pus,Wor3} and quantum algebras \cite{Dri1,Dri2,Jim1,Jim2}
 was applied to several fields of
theoretical physics. In particular, the one-parameter quantum algebra 
$U_{q}({\rm su}_2)$ was introduced in nuclear physics 
at the beginning of the 90's. In this direction, 
the $q$-rotator models developed in \cite{Iwa} (see also
\cite{Cap}) and in \cite{Ray} (see also \cite{Bon1,Bon2,Min}) 
are based on the $U_{q}({\rm su}_2)$
symmetry. In addition, the $\kappa$-Poincar\'e rotator model 
\cite{Cel} may be considered as a relativistic alternative to the $q$-rotator
models. 

In some preliminaries works, the authors showed the interest 
of using a two-parameter deformation of the Lie algebra u$_2$
for nuclear spectroscopy \cite{Bar1,Bar2} and molecular spectroscopy 
\cite{Bar3}. 
Along these lines, the junior author (R.~B.) studied in his thesis \cite{Bar4} 
the two-parameter quantum algebra $U_{qp}({\rm u}_2)$ and explored the
application of such an algebra to rotational collective dynamics in nuclei. 

The aim of this paper is two-fold: (i) to present the so-called $qp$-rotator 
model based on the $U_{qp}({\rm u}_2)$ symmetry and (ii) to report the 
main results, derived in \cite{Bar4},  
of its application to some 
superdeformed nuclei in the $A \approx 130$, $150$, and $190$ 
mass regions as well 
as to rare earth and actinide deformed nuclei.

   \vskip 1.2 true cm 

   {\bf 1. The $U_{qp}({\rm u}_2)$ quantum algebra}

   \vskip 0.6 true cm 

Two-parameter deformations of the algebra u$_2$ were worked out 
by several authors
(see, for instance, \cite{Dob,Kib,Cha}). We follow here Ref.~\cite{Kib} where
four generators $J_{\alpha}$ (with $\alpha = -, 0, +, 3$) span a deformation of
u$_2$ characterized by the following commutation relations  
 \begin{eqnarray}
        [J_0,J_\alpha ] & = &  0,                         \label{eq:1.2com1a} \\
\mbox{ }[J_{3},J_{\pm}] & = &  \pm J_{\pm},               \label{eq:1.2com1b} \\
\mbox{ }[J_{+}, J_{-} ] & = &  (qp)^{J_0-J_3} \ \qpn{2J_3},
                                                          \label{eq:1.2com1c}
 \end{eqnarray}
with the notation 
 \begin{equation}
 \qpn{X} := \frac{q^X - p^X}{q-p}. 
 \label{eq:1.2qpdef1}
 \end{equation}
The parameters $q$ and $p$ in (3) and (4) are {\it a priori} complexe
parameters. Hermiticity condition requirements show that they are allowed to
vary on two domains:
(i)  $q \in \gr$  and $p \in \gr$
and 
(ii) $q \in \grc$ and $p = \bar q \in \grc$,
where we exclude the values for which $q$ and $p$ are roots of unity. 
Note that by introducing the parameters $Q$ and $P$ defined by
 \begin{equation}
 Q := (qp^{-1})^{1\over 2}, \qquad 
 P := (qp)     ^{1\over 2}, 
 \label{eq:1.2para}
 \end{equation}
equation (3) can be rewritten as 
  \begin{equation}
  [J_{+}, J_{-} ]  =   P^{2J_0-1} \; \Qn{2J_3}, 
  \label{eq:1.2com2}
  \end{equation}
with   
  \begin{equation}
  \Qn{X} := \frac{Q^X - Q^{-X}}{Q-Q^{-1}}. 
  \label{eq:1.2qpdef1bis}
  \end{equation}
In the $QP$-parametrization, the domains (i) and (ii) corresponds to: 
(i) $(Q,P) \in \gr \times \gr$ and (ii) $(Q,P) \in {\rm S}^1 \times \gr$. 

The deformation spanned by the operators $J_{\alpha}$'s may be equipped 
with a Hopf algebraic structure \cite{Kib}, thus producing a quantum 
algebra denoted as $U_{qp}({\rm u}_2)$. Let us just mention that we can 
define a family of coproducts $\Delta^{QP}_{\beta}$ via 
 \begin{eqnarray}
 \Delta^{QP}_\beta(J_0) &:=& J_0  \otimes  {\bf 1} + {\bf 1}  \otimes  J_0, 
 \label{eq:1.2cop1b} \\
 \Delta^{QP}_\beta(J_3) &:=& J_3  \otimes  {\bf 1} + {\bf 1}  \otimes  J_3,
 \label{eq:1.2cop2b} \\
 \Delta^{QP}_\beta(J_+) &:=& J_{+}  \otimes P^{   \beta J_0}  Q^{+J_3} + 
 P^{(2-\beta)J_0} Q^{-J_3}   \otimes  J_{+},
 \label{eq:1.2cop3b} \\
 \Delta^{QP}_\beta(J_-) &:=& J_{-}  \otimes P^{(2-\beta)J_0}  Q^{+J_3} +
 P^{   \beta J_0}  Q^{-J_3}  \otimes  J_{-},
 \label{eq:1.2cop4b}
 \end{eqnarray}
where $\beta$ is a real parameter. Two extreme cases have been 
studied, namely, the cases corresponding to 
$\beta=1$ \cite{Kib} and 
$\beta=2$ \cite{Bar4}. 
For the latter two cases, we have the properties
 \begin{equation}
 \left(\Delta_1^{QP}(J_\pm) \right)^{\dagger} = \Delta_1^{QP}(J_\mp), 
 \qquad
 \left(\Delta_2^{QP}(J_\pm) \right)^{\dagger} = 
 \sigma \Delta_2^{Q^{-1}P}(J_\mp) 
 \label{eq:1.2adjo1} 
 \end{equation}
for the domain (i) and 
 \begin{equation}
 \left(\Delta_1^{QP}(J_\pm) \right)^{\dagger} = \Delta_1^{Q^{-1}P}(J_\mp),
 \qquad
 \left(\Delta_2^{QP}(J_\pm) \right)^{\dagger} = 
 \sigma \Delta_2^{QP}(J_\mp) 
 \label{eq:1.2adjo2} 
 \end{equation}
for the domain (ii). In (12) and (13), 
$X^{\dagger}$ stands for the adjoint of $X$ and 
$\sigma$ is the twist operator 
(such that $\sigma (a \otimes b) = b \otimes a$).
It is to be noted that for $p=q^{-1}$ (i.e., $P=1$), the quantum
algebra $U_{qp}({\rm u}_2)$ reduces to $U_{q}({\rm su}_2) \times {\rm u}_1$, 
where $U_{q}({\rm su}_2)$ is the {\it classical} one-parameter quantum algebra 
used in Refs.~\cite{Iwa,Ray} (see also \cite{Smi,Kas}). 

The operator 
  \begin{equation}
C_2(U_{qp}({\rm u}_2)) := 
\udm (J_+J_- + J_-J_+)\
+\ \udm \> \Qn{2} \> P^{2J_0-1} \;
\left( \Qn{J_3} \right)^2
 \label{eq:1.2cas1}
 \end{equation}
can be shown to be an invariant operator 
for the algebra $U_{qp}({\rm u}_2)$. The irreducible
representations of $U_{qp}({\rm u}_2)$ may be labelled by doublets 
$(j_0,j)$ with $j_0 \in \grz$ and $2j \in \grn$. The representation $(j_0,j)$
is spanned by the set $\{ |j_0, j, m \rangle~:~m=-j, -j+1, \cdots, +j \}$, where
the state vector $|j_0, j, m \rangle$ is obtained from the highest weight
state vector $|j_0, j, j \rangle$ through
  \begin{equation}
  \vert j_0, j, m \rangle \; = \; (qp)^{-{1 \over 2}(j_0-j)(j-m)}
  \, \sqrt {{ \qpn{j+m}! \over \qpn{2j}! \qpn{j-m}! }}
  \, (J_-)^{j-m} 
  \, \vert j_0, j, j \rangle, 
  \label{eq:kkaass}
  \end{equation}
where $\qpn{n}!$ is the $qp$-deformed factorial of $n \in \grn$. 
The eigenvalues of $C_2(U_{qp}({\rm u}_2))$ 
in the irreducible representation $(j_0,j)$ assume the form 
\begin{equation}
 (qp)^{j_0-j} \; \qpn{j} \; \qpn{j+1} \equiv P^{2j_0-1} \; \Qn{j} \; \Qn{j+1}
\label{eq:2.1eigen4}
\end{equation} 
and clearly depend on the two parameters $q$ and $p$ (or, alternatively, $Q$
and $P$). We have calculated the Clebsch-Gordan coefficients 
$\cg{j_{01}}{j_{02}}{j_1}{j_2}{m_1}{m_2}{j_0}{j}{m}_{\beta,qp}$ 
corresponding to the coproducts $\Delta_1$ (with $\beta = 1$)
                            and $\Delta_2$ (with $\beta = 2$). As a result, we
have
\begin{eqnarray}
\cg{j_{01}}{j_{02}}{j_1}{j_2}{m_1}{m_2}{j_0}{j}{m}_{1,qp}  & = & 
\delta_{j_0, j_{01} + j_{02}} \cgn{j_1}{j_2}{m_1}{m_2}{j}{m}_Q, 
\label{eq:2.2resu2} \\
\cg{j_{01}}{j_{02}}{j_1}{j_2}{m_1}{m_2}{j_0}{j}{m}_{2,qp}  & = &  
\delta_{j_0, j_{01} + j_{02}} \cgn{j_1}{j_2}{m_1}{m_2}{j}{m}_Q 
P^{j_{01}(j-m) - j_0(j_1-m_1)}, 
\label{eq:bbbiii}
\end{eqnarray}
where $\cgn{j_1}{j_2}{m_1}{m_2}{j}{m}_Q$ is the one-parameter Clebsch-Gordan
coefficient for the algebra $U_{q}({\rm su}_2)$ (see \cite{Smi} and \cite{Kas}).

   \vskip 1.2 true cm 

   {\bf 2. The $qp$-rotator model}

   \vskip 0.6 true cm 

Let us now list the basic hypotheses of the $qp$-rotator model for describing
rotational bands of nuclei. (The model is also of interest for diatomic
molecules.) As a first hypothesis, we take the Hamiltonian
\begin{equation}
H := { 1 \over 2{\cal J} } \; C_2 (U_{qp}({\rm u}_2)) + E_0,  
\label{eq:4.1ham1a}
\end{equation}
where the constants $E_0$ and ${\cal J}$ stand for 
the bandhead energy and the effective moment of inertia of a given nucleus, 
respectively. Such an Hamiltonian exhibits the 
 $U_{qp}({\rm u}_2)$ symmetry. As a second hypothesis, the physical state
vectors are
chosen as basis vectors for the irreducible representation $(I,I)$ (i.e.,
$j_0=j=I$, where $I$ is the angular momentum 
of the nucleus under consideration). Therefore,
the eigenvalues of $H$ turn out to be 
\begin{equation}
  E(I)_{qp} = { 1 \over {2 {\cal J}} } \; 
             \qpn{I} \; \qpn{I+1} + E_0
            = { 1 \over {2 {\cal J}} } \; P^{2I-1} \;
             \Qn{I}  \; \Qn{I+1}  + E_0.
\label{eq:4.1prop1b}
\end{equation}
Finally, the third hypothesis concerns the calculation of the $B$(E2)
electric-quadrupole transition 
probability for the $\gamma$-transition $(K:I+2) \to (K:I)$ between
the levels $I+2$ and $I$ of the $K$ band. We assume that the 
$B$E(2) reduced transition probability is defined by 
 \begin{equation}
B({\rm E2}; KI_1  \to KI_2)_{\beta,qp} := \frac {5} {16 \pi}\; Q_0^2\;
  \bigg  \vert \cg{I_1}{I_2-I_1}{I_1}{2}{K}{0}{I_2}{I_2}{K}_{\beta,qp}
  \bigg \vert ^2, 
\label{eq:4.2be2n2}
\end{equation}
for the $U_{qp}({\rm u}_2)$ symmetry, where $Q_0$ is the intrinsic
electric-quadrupole moment. Equations (16) and (17) lead to 
 \begin{equation}
  B({\rm E2}; KI+2  \to KI)_{1,qp} = 
  B({\rm E2}; KI+2  \to KI)_{Q }
\label{eq:4.2be2n4}
\end{equation} 
and
 \begin{equation}
  B({\rm E2}; KI+2  \to KI)_{2,qp} = P^{-4K} \;
  B({\rm E2}; KI+2  \to KI)_{Q }
\label{eq:4.2be2n4bis}
\end{equation}  
for $\beta =1$ and $\beta=2$, respectively, where 
$B({\rm E2}; KI+2  \to KI)_{Q }$ is the $B$(E2) 
reduced transition probability for the
$q$-rotator model developed in Refs.~\cite{Ray,Bon1,Bon2,Min} on the basis of
the $U_{q}({\rm su}_2)$ symmetry. 

   \vskip 1.2 true cm 

   {\bf 3. Physical results and conclusions}

   \vskip 0.6 true cm 

As a first test of the $qp$-rotator model presented in Section 2, we fitted
$\gamma$-transitions 
on experimental data for rotational bands of superdeformed (SD) and deformed
(D) nuclei. We have chosen two ranges of variation, compatible with the above
mentioned domains (i) and (ii), for the parameters $q$ and $p$. 
They correspond to the parametrizations: 
 \begin{equation}
{\rm (i)} \qquad 
q = {\rm e}^{a+b}, \quad 
p = {\rm e}^{a-b}  \qquad {\rm and} \qquad
{\rm (ii)} \qquad 
q = {\rm e}^{a+{\rm i}b}, \quad 
p = {\rm e}^{a-{\rm i}b}.  
  \label{eq:4.1par4} 
 \end{equation}
The relevant energy formulas used in our fitting procedures are 
 \begin{equation}
 E(I)_{qp} \equiv
 E(a,b;\ I)_{\rm (i)}   = \frac{1}{2 {\cal J}} \;{\rm e}^{(2I-1) a} \;
 \frac{\sinh (I b) \; \sinh [(I+1) b]}{\sinh^2 b}
 + E_0  
 \label{eq:4.1prop3b1} 
 \end{equation}
and
 \begin{equation}
 E(I)_{qp} \equiv
 E(a,b;\ I)_{\rm (ii)}  = \frac{1}{2 {\cal J}} \;{\rm e}^{(2I-1) a} \;
 \frac{\sin (I b) \; \sin [(I+1) b]}{\sin^2 b}
 + E_0  
 \label{eq:4.1prop3b2} 
 \end{equation}
for the cases (i) and (ii), respectively. The fitted
values of the $qp$-rotator parameters 
(${\cal J}$, $a$, and $b$) were obtained by
minimizing the standard deviation 
    \begin{equation}
\chi := \sqrt{ { 1 \over {n -  m}} \; \; \sum_I \; \;
         \left[ { {E_{\gamma}^{\rm th } (I) -
                   E_{\gamma}^{\rm ex } (I)} \over
           {\Delta E_{\gamma}^{\rm ex } (I)} } \right]^2},
\label{eq:6.1qui}
\end{equation}
where $n$ is the number of fitted $\gamma$-transitions, $m$ the number of
fitting parameters, and $\Delta_{\gamma}^{\rm ex}(I)$ the experimental error 
for the $\gamma$-transition from the $E(I)_{qp}$ level to the $E(I-2)_{qp}$ 
level. 

As a second important test of our model, we calculated the
theoretical dynamical moment of inertia ${\cal J}_{{\rm th}}^{(2)}$ defined by
\begin{equation}
{\cal J}_{{\rm th}}^{(2)}(I) := 
\left( \frac { d ^2 E } { d x ^2 }  \right)^{-1},
\quad E \equiv E(I)_{qp},
\quad x \equiv x(I) := \sqrt{I(I+1)}.
\label{eq:7.1iner2}
\end{equation}
For each considered nucleus, the moments of inertia 
${\cal J}_{{\rm th}}^{(2)}$ were 
calculated, as a function of the angular momentum 
$I$, from the values obtained for the fitting parameters 
(${\cal J}$, $a$, and $b$) and compared to the experimental ones.  

The two tests were performed on 20 SD bands in the $A \approx 130$, 150, and 
190 mass regions and on 29 D bands in the rare earth and actinide mass regions.
For the purpose of comparison, the same tests 
were achieved through the use of the $q$-rotator models
of Refs.~\cite{Iwa} and \cite{Ray} and of the $\kappa$-Poincar\'e model 
of Ref.~\cite{Cel}. 
The main results of our analyses can be summed up as follows. 

(i) The results obtained from the $qp$-rotator model are better, both 
for the (fitted) $\gamma$-transitions and the (calculated) 
dynamical moments of inertia, than the ones derived from the $q$-rotator
models and the $\kappa$-Poincar\'e model (see \cite{Bar4} 
for an exhaustive study). 

(ii) The best results for the $qp$-rotator model are obtained with the first
domain of variation of the parameters $q$ and $p$ (i.e., with (25)) in the $A
\approx 130$ and 150 mass regions and with the second domain 
(i.e., with (26)) in the $A \approx 190$, rare earth, and actinide 
mass regions. It is to be noted that the latter fact parallels the 
experimental situation according to which 
the dynamical moments of inertia decrease (respectively, increase)
for the $A \approx 130$ and 150 SD bands (respectively, 
for the $A \approx 190$         SD bands and for the rare earth and actinide D
bands). Furthermore, it is also to be noted that the dynamical 
moments of inertia derived
from the $q$-rotator models and the $\kappa$-Poincar\'e model are not in good
agreement with the corresponding experimental values in the $A \approx 130$ and
150 mass regions. 

(iii) The ranges of variation of the $qp$-rotator parameters $a$ and
$b$ depend on the bands under consideration. Indeed, the parameter $b$
(respectively $a$) is 
of the order 10$^{-3}$ (respectively, 10$^{-4}$) for the 
$A \approx 190$ SD bands and 
of the order 10$^{-2}$ (respectively, 10$^{-3}$) for the
rare earth and actinide D bands. This is in accordance with the experimental
fact that the increasing of the dynamical moment of inertia is more important
in the D bands than in the $A \approx 190$ 
SD bands. In other words, the two parameters $a$ and
$b$ of the $qp$-rotator model can be interpreted as inertial parameters that
describe the softness of the D and SD nuclei. 

(iv) To close this paper, let us emphasize that the second deformation
parameter, viz., $p$, of the quantum algebra $U_{qp}({\rm u}_2)$
plays an important r\^ole via the factor ${\rm e}^a$. The parameter 
$a$ ($a$ is 10 times lower than $b$)
acts as a correction of the softness mainly described by the parameter $b$. 
This is especially evident when comparing our
results with the ones given by the  $q$-rotator model of Ref.~\cite{Ray}. 
Finally, the fact that the $qp$-rotator model yields better results than the
$\kappa$-Poincar\'e model for rare earth and actinide nuclei shows that our 
model works also well for heavy nuclei and thus take into account, in a
phenomenological way, some relativistic effects. 

\vskip 1.2 cm

\noindent {\bf Acknowledgments}

\vskip 0.6 cm

The authors would like to thank Profs.~J.~Katriel, J.~Meyer, 
and Yu.~F.~Smirnov for interesting comments on this work. This 
work was presented by one of the authors (R.~B.)
to the symposium ``Quantum Groups and Their Applications in Physics'' 
(Pozna\'n, Poland, 17-20 October 1995). Thanks are due to 
Prof.~T.~Lulek for making it possible to present this contribution, 
on a physical application of quantum algebras, to the symposium.  

\newpage


\end{document}